\documentclass[reprint,floatfix,showpacs,aps,prx,twocolumn,twoside]{revtex4-2}
\usepackage{makeidx}
\usepackage{eurosym}
\usepackage{graphicx}
\usepackage{graphics}
\usepackage{amssymb}
\usepackage{amsmath}
\usepackage{color}
\usepackage{float}
\usepackage{hyperref}
\usepackage{fancyhdr}
\usepackage{booktabs}

\begin{document}

\title{Fluctuations superconductivity and giant negative magnetoresistance
in a gate voltage tuned 2D electron liquid with strong spin-orbit impurity
scattering }
\author{Tsofar Maniv}
\email{maniv@tx.technion.ac.il}
\author{Vladimir Zhuravlev}
\affiliation{Schulich Faculty of Chemistry, Technion-Israel Institute of
	Technology, Haifa 32000, Israel}
\date{\today }

\begin{abstract}
We present a quantitative theory of the gate-voltage tuned
superconductor-to-insulator transition (SIT) observed experimentally in the
2D electron liquid created in the (111) interface between crystalline SrTiO$%
_{3}$ and LaAlO$_{3}$. Considering two fundamental opposing effects of
Cooper-pair fluctuations; the critical conductivity enhancement, known as
para-conductivity, and its suppression associated with the loss of unpaired
electrons due to Cooper-pairs formation, we employ the standard thermal
fluctuations theory, modified to include quantum fluctuations within a novel
phenomenological approach. Relying on the quantitative agreement found
between our theory and a large body of experimental sheet-resistance data,
we conclude that spin-orbit scatterings, via significant enhancement of the
interaction between fluctuations, strongly enhance the sheet resistance peak
at high fields, and reveal anomalous metallic behavior at low fields, due to
mixing of relatively heavy electron bands with a light electron band near a
Lifshitz point.
\end{abstract}

\maketitle

Recently, it has been shown \cite{Mograbi19} that the two-dimensional (2D)
electron liquid formed at the (111) interface between the two insulators:
SrTiO$_{3}$ and LaAlO$_{3}$, can be smoothly tuned by gate bias from the
superconducting (SC) state deep into an insulating state with pronounced
magnetoresistance (MR) peaks developed at low temperatures. Similar
electrostatically tuned superconductor-to-insulator transition (SIT) was
reported for the LaAlO$_{3}$/SrTiO$_{3}$ (001) interface \cite{Caviglia08},
showing however \cite{Biscaras13},\cite{Mehta14} no clear indication of MR
peaks similar to those reported for the (111) interface. Earlier studies of
the (111) interface have found coexistence of magnetism and 2D
superconductivity \cite{Davis17}, and a correlation between
superconductivity and strong spin-orbit interaction \cite{RoutPRL17}. The
linear magnetic field dependence observed at low perpendicular fields and
its hysteresis, have indicated the importance of flux-flow in the detected
resistance. These effects persisted deep into the insulating state,
revealing the importance of Cooper-pair fluctuations even when
superconductivity is completely suppressed. The large smearing of the SC
resistive transitions observed under parallel fields may also reflect strong
SC fluctuations effect. The transition temperature, $T_{c}$ and the critical
fields, $H_{c\Vert }$, $H_{c\bot }$ for both parallel and perpendicular
fields, respectively, were found \cite{RoutPRL17} to follow non-monotonic
(dome-shaped) gate-voltage dependence of the spin-orbit interaction.

The phenomenon of SIT has been investigated for many years, notably in thin
films of materials like bismuth \cite{Haviland89}, InO \cite{Hebard90}, MoGe 
\cite{Yazdany95}, TiN \cite{Baturina07}, and cuprate superconductors \cite%
{Bollinger11}. Many intriguing phenomena have been associated with the
observation of SIT. Noteworthy examples are: scaling behavior near a quantum
critical point \cite{Hebard90},\cite{Ovadia13},\cite{Steiner05}, large MR
peaks \cite{Samb04}, \cite{Steiner05}, \cite{Baturina07} and thermally
activated insulating behavior \cite{Samb04}, \cite{Steiner05}, \cite%
{Baturina07}, \cite{Stewart07}. However, some of these effects have not been
observed in all materials that exhibit a SIT, making the interpretation of
the various SIT phenomena controversial, with no consensus as to their
mechanism and expected behavior.

Here we present a scenario of SIT in a 2D electron system, based on the
opposing effects generated by fluctuations in the SC order parameter: On the
one hand, the singular enhancement of conductivity due to fluctuating Cooper
pairs in approaching the critical magnetic field (paraconductivity), and on
the other hand, the suppression of conductivity associated with the loss of
unpaired electrons resulting from Cooper pairs formation. Specializing this
scenario to the LaAlO$_{3}$/SrTiO$_{3}$ (111) interface, the strongly
enhanced fluctuations effect is due to the remarkable combination of
marginal superconductivity driven by spin-orbit scattering versus the
pair-breaking of Zeeman spin-splitting effect embodied in a 2D electron
system. Focusing on the parallel field orientation case enables us to
investigate the essence of our model of SIT without interference from the
complex vortex kinetics and flux lines pinning processes involved in the
perpendicular field case. Furthermore, the striking observations of giant
negative MR in both the (111) and the (001) LaAlO$_{3}$/SrTiO$_{3}$
interfaces, driven by spin-orbit coupling under parallel field, at
temperatures above the SC transition \cite{RoutPRB17}, \cite{DiezPRL15}, are
of special interest here: It has been associated \cite{DiezPRL15} with
spin-orbit induced band mixing between orbitals of different symmetries near
a Lifshitz point in the d-electron interface band structure \cite%
{JoshuaNcomm12}.

We test the validity of the above mentioned SIT scenario by performing
calculations based on an effective mixed-bands DOS model, and comparing the
results with the large body of experimental magnetic sheet resistance (MSR)
data presented in Ref.\cite{Mograbi19}. The calculations were done within an
extended version of the Fulde-Maki Aslamazov-Larkin theory of fluctuations
in paramagnetically-limited superconductors \cite{FuldeMaki70},\cite%
{AoiPRB74},\cite{AL68}, in which the linear time-dependent-Ginzburg-Landau
(TDGL) equation describing free (Gaussian) fluctuations is modified by
taking into account interactions between free fluctuations,
self-consistently in the Hartree approximation \cite{UllDor90}. Very good
quantitative agreement between theory and experiment has been achieved,
confirming our SIT scenario. Dynamical quantum tunneling of cooper-pair
fluctuations through GL energy barriers is taken into account, on equal
footing with thermal activation, within a phenomenological approach,
preserving the level of agreement with the experiment in the low
temperatures region. Our calculations also reveal how a Lifshitz transition
in the d-electron interface band structure \cite{KhannaPRL19}, \cite%
{JoshuaNcomm12} can drive large enhancements of the MSR peak observed at the
end of the SIT path \cite{Mograbi19} upon gate-voltage variation. This
feature is exploited to construct the mixed-band DOS function around the
Lifshitz-point from the gate-voltage dependence MSR data.

We consider a planar, 2D SC electron system subject to strong spin-orbit
impurity scattering \cite{Abrikosov62}, \cite{Klemm75}, with a
characteristic relaxation time $\tau _{SO}=\hbar /\varepsilon _{SO}$,\ under
a strong magnetic field $H$, applied parallel to the plane.
Superconductivity in this system is governed both by the Zeeman spin
splitting energy,$\mu _{B}H$ and the spin-orbit scattering rate, $1/\tau
_{SO}$ (see a detailed description in Supplemental Material (SM) \cite%
{SuppMat} and in early papers dealing with similar 3D systems \cite{Maki66},%
\cite{FuldeMaki70},\cite{WHH66}). The underlying spin-orbit induced band
mixing, which was evaluated microscopically for the (001) interface by
several authors \cite{JoshuaNcomm12}, \cite{Ruhman2014}, \cite{DiezPRL15}, 
\cite{BovenziPRB17}, is taken into account here phenomenologically within a
minimal model of 2D DOS function $N_{2D}\left( E\right) $ (see Fig.(1)).
Consistently with this, we use in our transport calculations a 2D electon
gas, in the dirty limit, confined within a thin rectangular film of
thickness $d$. In deriving the corresponding self-consistent field (SCF)
equation (see SM \cite{SuppMat}) a key quantity is the Cooper-pair
fluctuation propagator: $D\left( q;\varepsilon _{H}\right) =1/N_{2D}\left(
E_{F}\right) \Phi \left( x;\varepsilon _{H}\right) $, obtained from the
well-known function of the dimensionless fluctuation kinetic energy variable 
$x\equiv \hbar Dq^{2}/4\pi k_{B}T$\cite{FuldeMaki70}:

\begin{eqnarray}
\Phi \left( x;\varepsilon _{H}\right) &=&\varepsilon _{H}+a_{+}\left[ \psi
\left( 1/2+f_{-}+x\right) -\psi \left( 1/2+f_{-}\right) \right]  \notag \\
&&+a_{-}\left[ \psi \left( 1/2+f_{+}+x\right) -\psi \left( 1/2+f_{+}\right) %
\right]  \label{Phi(x)}
\end{eqnarray}%
and the Gaussian critical shift-parameter: 
\begin{equation}
\varepsilon _{H}\equiv \ln \left( \frac{T}{T_{c0}}\right) +a_{+}\psi \left( 
\frac{1}{2}+f_{-}\right) +a_{-}\psi \left( \frac{1}{2}+f_{+}\right) -\psi
\left( 1/2\right)  \label{eps_H}
\end{equation}

Here $T_{c0}$ is the mean-field SC transition temperature at zero magnetic
field, $\psi $ is the digamma function, $\ f_{\pm }=\delta H^{2}+\beta \pm 
\sqrt{\beta ^{2}-\mu ^{2}H^{2}}$, $a_{\pm }=\left( 1\pm \beta /\sqrt{\beta
^{2}-\mu ^{2}H^{2}}\right) /2$ are functions of the magnetic field $H$, with
the basic parameters: $\  \beta \equiv \varepsilon _{SO}/4\pi k_{B}T,$ $\mu
\equiv \mu _{B}/2\pi k_{B}T,$ $\delta \equiv D\left( de\right) ^{2}/2\pi
k_{B}T\hslash $, where $\mu _{B}=e\hbar /2m_{e}$ is the Bohr magneton, $%
D\equiv \hbar E_{F}/m^{\ast }\varepsilon _{SO}$ the electron diffusion
coefficient, $E_{F}$ the Fermi energy and $m^{\ast }$is the band effective
mass.

\begin{figure}[tbp]
\label{fig1} \includegraphics[width=3.3in]{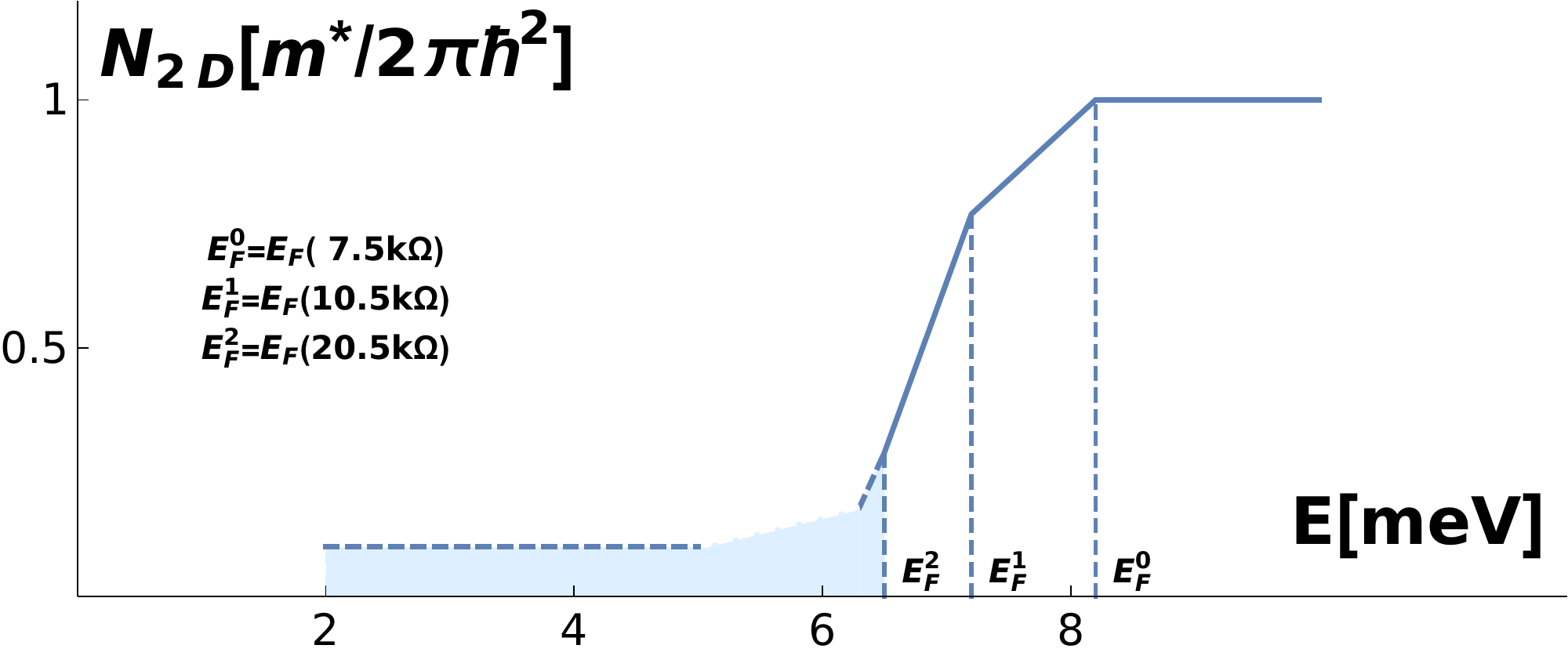}
\caption{A three-point histogram of the 2D DOS function extracted in our
fitting process from the sheet resistance data for three values of $R_{N}$.
The corresponding values of the interaction parameters are: $\protect \alpha %
\left( 7.5k\Omega \right) =0.09$, $\protect \alpha \left( 10.5k\Omega \right)
=0.11$,and $\protect \alpha \left( 20.5k\Omega \right) =0.30.$ }
\end{figure}

The values of the fluctuation wavenumber $q$ are bound by a cutoff $q_{c}$,
which typically satisfies: $x_{c}\equiv \hbar Dq_{c}^{2}/4\pi k_{B}T<1$, so
that one may exploit the linear approximation: $\Phi \left( x;\varepsilon
_{H}\right) =\varepsilon _{H}+\eta \left( H\right) x$. The Gaussian critical
shift parameter $\varepsilon _{H}$ (Eq.\ref{eps_H}) should be corrected, due
to interaction between fluctuations \cite{UllDor91}. The correction can be
evaluated analytically from the cubic term of the GL equation discussed
above, and is given by (see SM \cite{SuppMat}): $\alpha F\left( H\right)
\eta \left( H\right) \int_{0}^{x_{c}}dx/\Phi \left( x;\varepsilon
_{H}\right) $, where: 
\begin{equation}
\alpha \equiv 1/\hslash \pi ^{3}DN_{2D}\left( E_{F}\right)  \label{alpha}
\end{equation}%
The Hartree SCF approximation amounts to replacing $\varepsilon _{H}$,
appearing in the the interaction correction, with the "dressed" critical
shift-parameter $\widetilde{\varepsilon }_{H}$, leading to the SCF equation:

\begin{equation}
\widetilde{\varepsilon }_{H}\simeq \varepsilon _{H}+\alpha F\left( H\right)
\ln \left( 1+\frac{\eta \left( H\right) x_{c}}{\widetilde{\varepsilon }_{H}}%
\right)  \label{SCFeq}
\end{equation}%
where the logarithmic factor is obtained from the integral over $x$ by using
the linear approximation of $\Phi \left( x;\widetilde{\varepsilon }%
_{H}\right) $, and the field distribution function of the interaction $%
F\left( H\right) $, is given by the Matzubara sum \cite{SuppMat}:

\begin{equation}
F\left( H\right) =\frac{1}{\eta \left( H\right) }\sum \limits_{n=0}^{\infty }%
\frac{\varkappa _{n}\left( \varkappa _{n}^{2}+\mu ^{2}H^{2}\right) }{\left[
\varkappa _{n}\left( \varkappa _{n}-2\beta \right) +\mu ^{2}H^{2}\right] ^{3}%
}  \label{F}
\end{equation}%
where $\varkappa _{n}=n+1/2+2\beta +\delta H^{2}$. \ Equation (\ref{SCFeq})
has no solution with $\widetilde{\varepsilon }\leq 0$ (see Ref.\cite%
{UllDor91}), indicating the absence of a genuine SC phase transition due to
the interaction between fluctuations. Indeed, as shown in Fig.(2), all
solutions of the SCF equation \ref{SCFeq} satisfy $\widetilde{\varepsilon }%
_{h}>0$, implying that the critical divergence of the free fluctuations
propagator is strictly removed. This also eliminates the critical divergence
from both the Aslamazov-Larkin (AL) and the suppressed normal-state
conductivities (see below).

\begin{figure}[tbp]
\label{fig2} \includegraphics[width=3.2in]{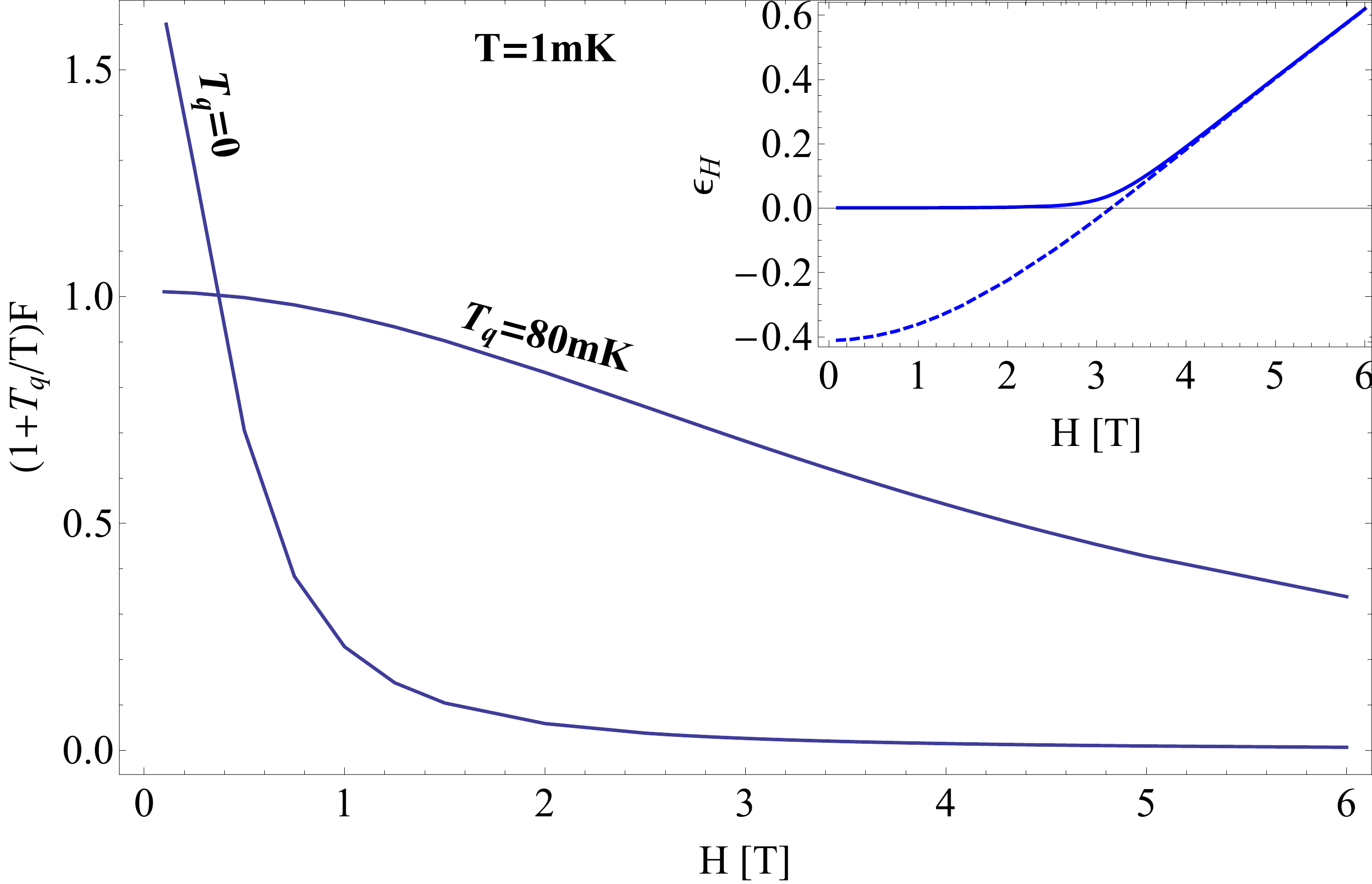}
\caption{The field distribution function $\left( 1+T_{Q}/T\right)
F_{U}\left( H\right) $, calculated at $T=1$mK for $T_{Q}=0$ and $T_{Q}=80$
mK. Inset: "Bare", $\protect \varepsilon _{H}$ (dashed line), and "dressed", $%
\widetilde{\protect \varepsilon }_{H}$ (solid line), critical shift
parameters, calculated at $T=1$mK and $T_{Q}=80$mK.}
\end{figure}

\textit{Quantum fluctuations at low temperatures--}For temperatures above $%
T_{c}$, our calculated MSR accounts quantitatively well for the experimental
MSR data reported in \cite{Mograbi19}. However, in the low temperatures
regime well below $T_{c}$, large deviations between the calculated and
measured MSR data were found (see SM \cite{SuppMat}), with the calculated
MSR peak quickly narrowing upon decreasing temperature, as compared to the
rather broad experimental MSR peak. This discrepancy is due to the fact
that, at low temperatures $\widetilde{\varepsilon }_{H}$, determined by Eq.%
\ref{SCFeq}, is not significantly different from $\varepsilon _{H}$ in the
vicinity of the critical point $\varepsilon _{H}=0$. The reason, as
illustrated by Fig.(2), is in the progressive narrowing of $F\left( H\right) 
$ (Eq.\ref{F}) upon decreasing temperature, having too small tail intensity
in the vicinity of the free-fluctuations critical field.

We argue that the observed broadening at very low temperatures is due to
quantum fluctuations effect similar to the quantum phase slips reported for
SC nanowires \cite{Tinkham2000},\cite{Mooij2006}, \cite{Arutyunov-etal2008}, 
\cite{Lehtinen2012}, a phenomenon which was also reported for ultrathin
granular SC film \cite{Goldman86}, \cite{Fisher86}. We invoke a
phenomenological approach describing dynamical tunneling of Cooper-pairs
through energy barriers, separating SC puddles \cite{HurandPRB19}, on equal
footing with thermal activation across the same barriers. Thus, we introduce
a unified quantum-thermal (QT) fluctuations partition function: 
\begin{equation*}
Z_{fluct}^{U}=\prod \limits_{\mathbf{q}}\int \mathcal{D}\Delta _{q}\mathcal{D%
}\Delta _{q}^{\ast }e^{-\frac{\tau _{U}}{\hbar }\left[ \widetilde{%
\varepsilon }_{H}^{U}+\frac{\eta _{U}\left( H\right) \hbar }{4\pi k_{B}T}%
Dq^{2}\right] \left \vert \Delta _{q}\right \vert ^{2}N_{2D}\left(
E_{F}\right) },
\end{equation*}%
where $1/\tau _{U}$ is the combined QT attempt rate, defined by: $\hbar
/\tau _{U}\equiv k_{B}T+\hbar /\tau _{Q}$, with $1/\tau
_{Q}=k_{B}T_{Q}/\hbar $ the tunneling attempt rate ($\widetilde{\varepsilon }%
_{H}^{U}$, and $\eta _{U}\left( H\right) $ will be defined below). The
corresponding Gaussian, unified QT-fluctuations propagator is given by: $%
D_{U}\left( q;\varepsilon _{H}^{U}\right) =k_{B}\left( T+T_{Q}\right)
/N_{2D}\left( E_{F}\right) \left( \varepsilon _{H}^{U}+\frac{\hbar
Dq^{2}\eta _{U}\left( H\right) }{4\pi k_{B}T}\right) $.

The inherent dynamics of the quantum tunneling of Cooper-pair fluctuations
is introduced to the equilibrium Gorkov-GL functional-integral through
imaginary time \cite{Hertz76}. Consistency requires that the introduction of
an excess quantum-tunneling "temperature", $T_{Q}$, into the unified QT
fluctuation propagator, should be complemented by introduction of a bosonic
excitation Matzubara frequency-shift $\Omega _{\nu }/2=\pi
k_{B}T_{Q}/\hslash $ into the definitions of the electron-pairing functions $%
F_{U}\left( H\right) ,\varepsilon _{H}^{U},\eta _{U}\left( H\right) $, under
summation over the fermionic Matzubara frequency $\omega _{n}=\left(
2n+1\right) \pi k_{B}T/\hslash $. Thus, one evaluates these unified QT
functions from the respective thermal functions: $F\left( H\right)
,\varepsilon _{H},\eta \left( H\right) $, by introducing the shift $%
n\rightarrow n\mathbf{+}T_{Q}/2T$ under the summations over $n$ in Eq.\ref{F}%
, and by shifting the argument of the digamma function and its derivative,
respectively, with the same additive constant $T_{Q}/2T$ in Eqs.\ref{Phi(x)}
and \ref{eps_H}.\ The corresponding unified Hartree SCF equation \ref{SCFeq}
reads: $\widetilde{\varepsilon }_{H}^{U}=\varepsilon _{H}^{U}+\alpha
F_{U}\left( H\right) \left( 1+T_{Q}/T\right) \ln \left( 1+\frac{x_{c}\eta
_{U}\left( H\right) }{\widetilde{\varepsilon }_{H}^{U}}\right) $.

\textit{Fluctuation paraconductivity under a parallel magnetic field--}In
the calculation of the paraconductivity we adopt a modified version of the
formalism developed by Fulde and Maki \cite{FuldeMaki70} for calculating the
AL contribution \cite{AL68}. Thus, the static AL sheet conductivity,
calculated in the unified QT fluctuations approach (see SM \cite{SuppMat})
is given by: 
\begin{equation}
\sigma _{AL}^{U}d=\left( 1+\frac{T_{Q}}{T}\right) \frac{1}{4}\left( \frac{%
G_{0}}{\pi }\right) \int \limits_{0}^{\infty }\left( \frac{\Phi _{U}^{\prime
}\left( x;\widetilde{\varepsilon }_{H}^{U}\right) }{\Phi _{U}\left( x;%
\widetilde{\varepsilon }_{H}^{U}\right) }\right) ^{2}dx  \label{sigAL^Ud}
\end{equation}%
where $G_{0}=e^{2}/\pi \hbar $ is the conductance quantum, and $\Phi
_{U}\left( x;\widetilde{\varepsilon }_{H}^{U}\right) $ is obtained from Eq.%
\ref{Phi(x)} by replacing $\widetilde{\varepsilon }_{H}$ with $\widetilde{%
\varepsilon }_{H}^{U}$, and by shifting the argument of all the digamma
functions in Eq.\ref{Phi(x)} with the additive constant $T_{Q}/2T$. \ 

\textit{Cooper-pair fluctuations suppressed normal state conductivity--}The
idea, first exploited by Larkin and Varlamov \cite{LV05} for the zero field
case, is to replace the electron number density $N_{e}$ in the simple Drude
formula for the conductivity $\sigma =N_{e}e^{2}\tau /m^{\ast }$, with the
number density of electrons occupying quasi-particle states minus the number
density $\Delta N_{e}$ of electrons paired into SC puddles. Since $\Delta
N_{e}=2n_{s}$, where $n_{s}$ is the number density of Cooper pairs in SC
puddles, the corresponding correction to the Drude conductivity is given by: 
$\delta \sigma _{DOS}=-2\left( n_{s}e^{2}/m^{\ast }\right) \tau _{SO}$. The
subscript DOS indicates that this contribution to the conductivity is
associated with the suppression of the normal electrons DOS by Cooper-pair
fluctuations. The number density, $n_{s}=\left( 1/d\right) \int
\left
\langle \left \vert \psi \left( q\right) \right \vert
^{2}\right
\rangle d^{2}q/\left( 2\pi \right) ^{2}$, \cite{LV05} is
obtained from the superfluid momentum distribution-function $\left \langle
\left \vert \psi \left( q\right) \right \vert ^{2}\right \rangle \simeq
2E_{F}/\pi ^{2}k_{B}T\Phi \left( x;\widetilde{\varepsilon }_{H}\right) $, so
that: $\delta \sigma _{DOS}d\simeq -4\left( G_{0}/\pi \right)
\int_{0}^{x_{c}}dx/\Phi \left( x;\widetilde{\varepsilon }_{H}\right) $. This
result is, to a good approximation, equal to the DOS conductivity obtained
in \cite{LV05} by means of a microscopic (diagrammatic) approach in the
dirty limit (see SM \cite{SuppMat}). The unified QT fluctuations version of
the DOS conductivity can be derived by introducing quantum fluctuations into
the superfluid momentum distribution function as follows: $2E_{F}/\pi
^{2}k_{B}T\Phi \left( x;\widetilde{\varepsilon }_{H}\right) \rightarrow
2E_{F}/\pi ^{2}k_{B}\left( T+T_{Q}\right) \Phi _{U}\left( x;\widetilde{%
\varepsilon }_{H}^{U}\right) $, resulting in the following expression:

\begin{equation}
\delta \sigma _{DOS}^{U}d\simeq -4\left( \frac{G_{0}}{\pi }\right)
\int_{0}^{x_{c}}\frac{dx}{\left( 1+\frac{T_{Q}}{T}\right) \Phi _{U}\left( x;%
\widetilde{\varepsilon }_{H}^{U}\right) }  \label{delsig_DOS^Ud}
\end{equation}

It is shown in SM (\cite{SuppMat}) that both $\sigma _{AL}^{U}$ and $\delta
\sigma _{DOS}^{U}$ have well defined quantum ($T\rightarrow 0$) limit.

Combining all contributions to the sheet conductivity, Eqs.\ref{sigAL^Ud},%
\ref{delsig_DOS^Ud}, including the normal-state conductivity $\sigma _{n}$,
we have:

\begin{equation}
\sigma ^{U}d=\sigma _{n}d+\sigma _{AL}^{U}d+\delta \sigma _{DOS}^{U}d
\label{sig^Ud}
\end{equation}

Determination of the normal-state conductivity, $\sigma _{n}$, can reflect
on the strong negative MR reported in Ref.\cite{RoutPRB17} for temperatures
well above the SC transition. Thus, in our fitting procedure we assume a
field-dependent normal state conductivity contribution $\sigma _{n}\left(
H,T\right) $, which produces negative MR similar to that observed in Ref.%
\cite{RoutPRB17}, by employing the quadratic function: $\sigma _{n}\left(
H,T\right) =\sigma _{0}+\sigma _{0}\left( H/H_{n}\left( T\right) \right)
^{2} $, with two adjustable parameters $\sigma _{0},H_{n}\left( T\right) $,
where the latter is temperature dependent. Employing an extensive fitting
procedure, as described in detail in SM \cite{SuppMat}, the resulting
calculated MSR, best fit to the experimental data sets \cite{Mograbi19}, are
shown in Fig.(3). Very good quantitative agreement between the calculated
and measured data is seen for the entire data presented. The decreasing
magnitudes of the normal-state MR curves, shown in Fig.(3), with increasing
temperature are seen to be in qualitative agreement with the experimental
negative MR data, presented in Ref.\cite{RoutPRB17} for temperatures well
above $T_{c}$.

\begin{figure}[tbh!]
\label{fig3} \includegraphics[width=3.3in]{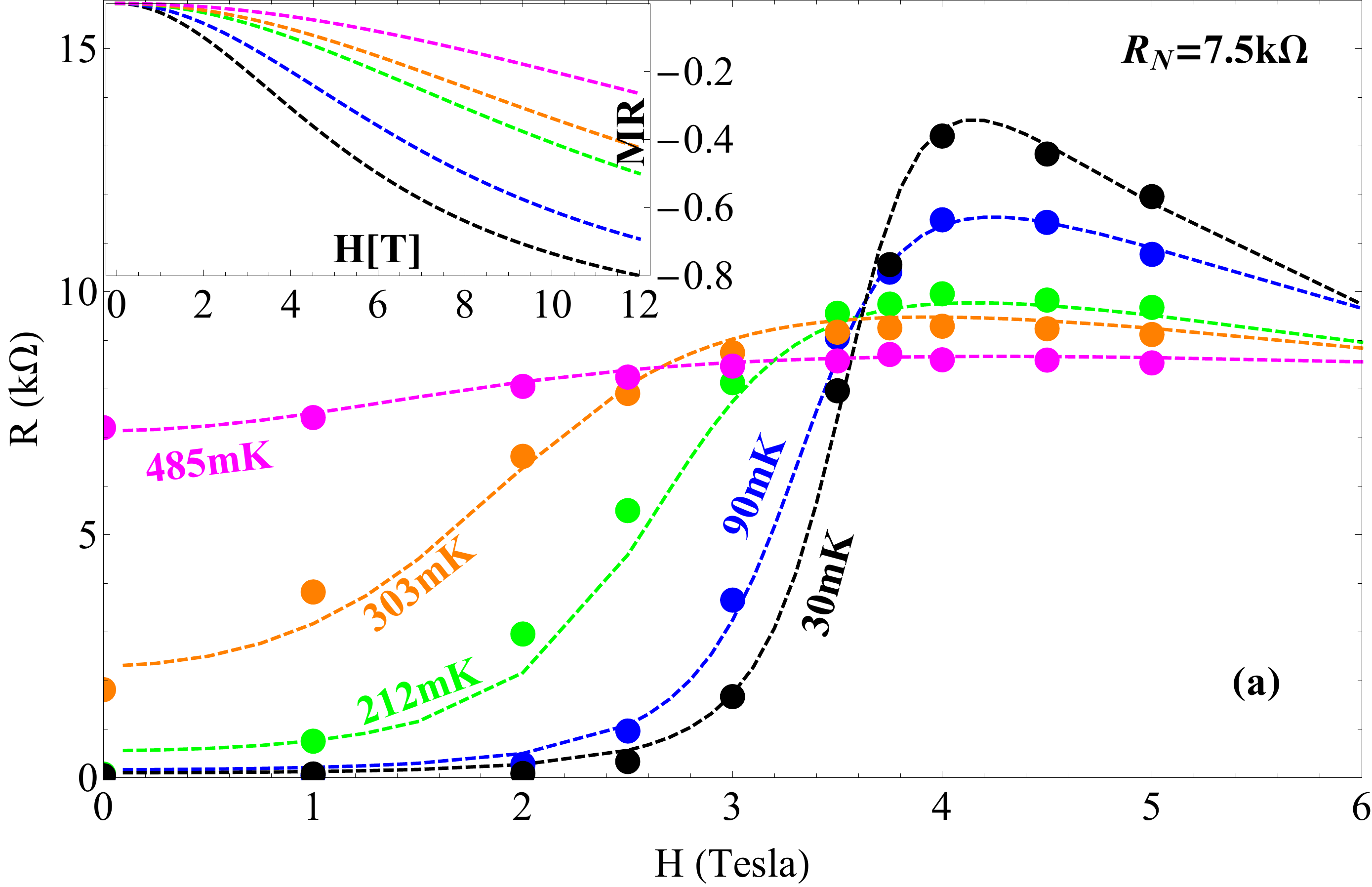} %
\includegraphics[width=3.3in]{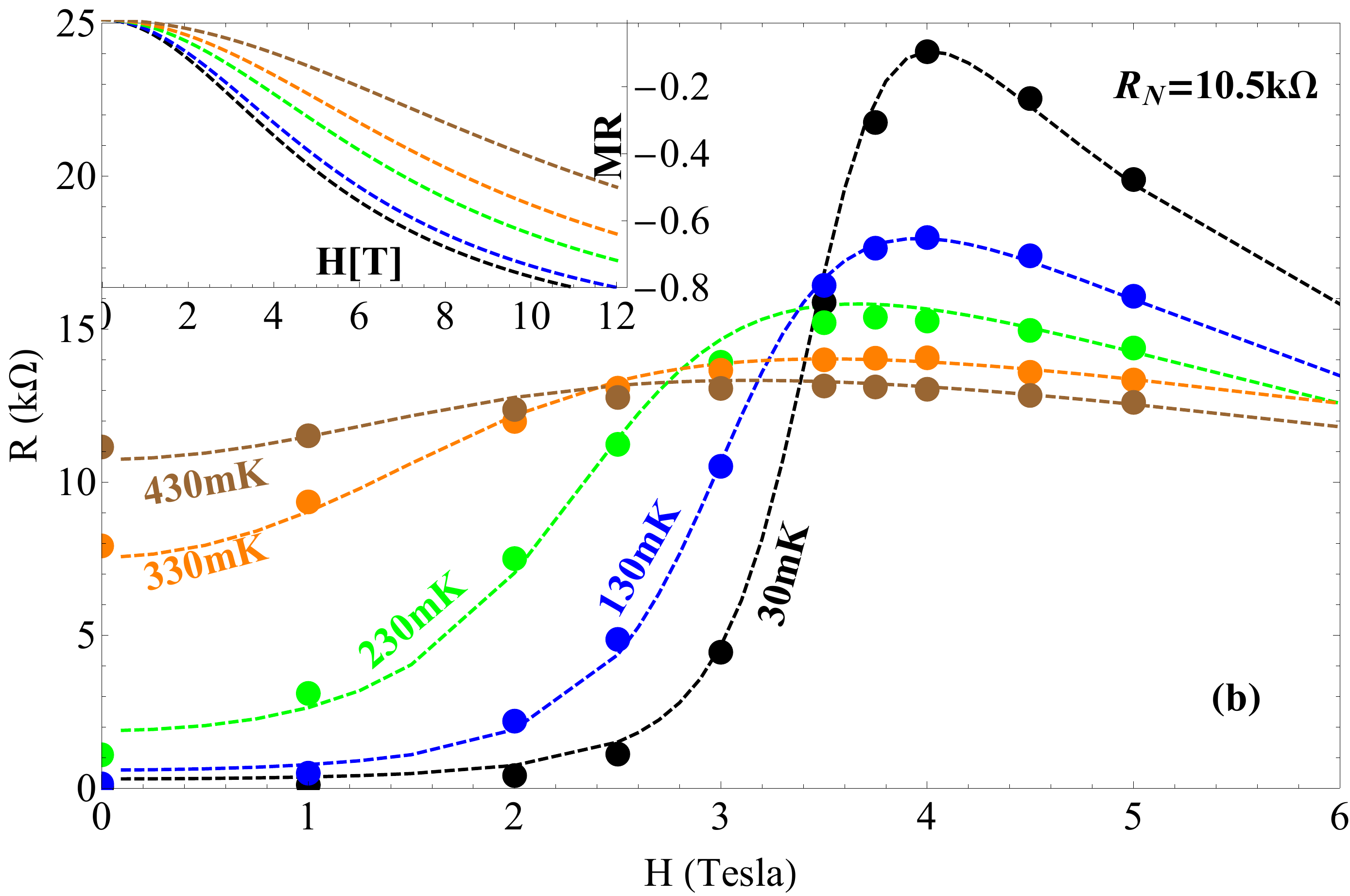} %
\includegraphics[width=3.3in]{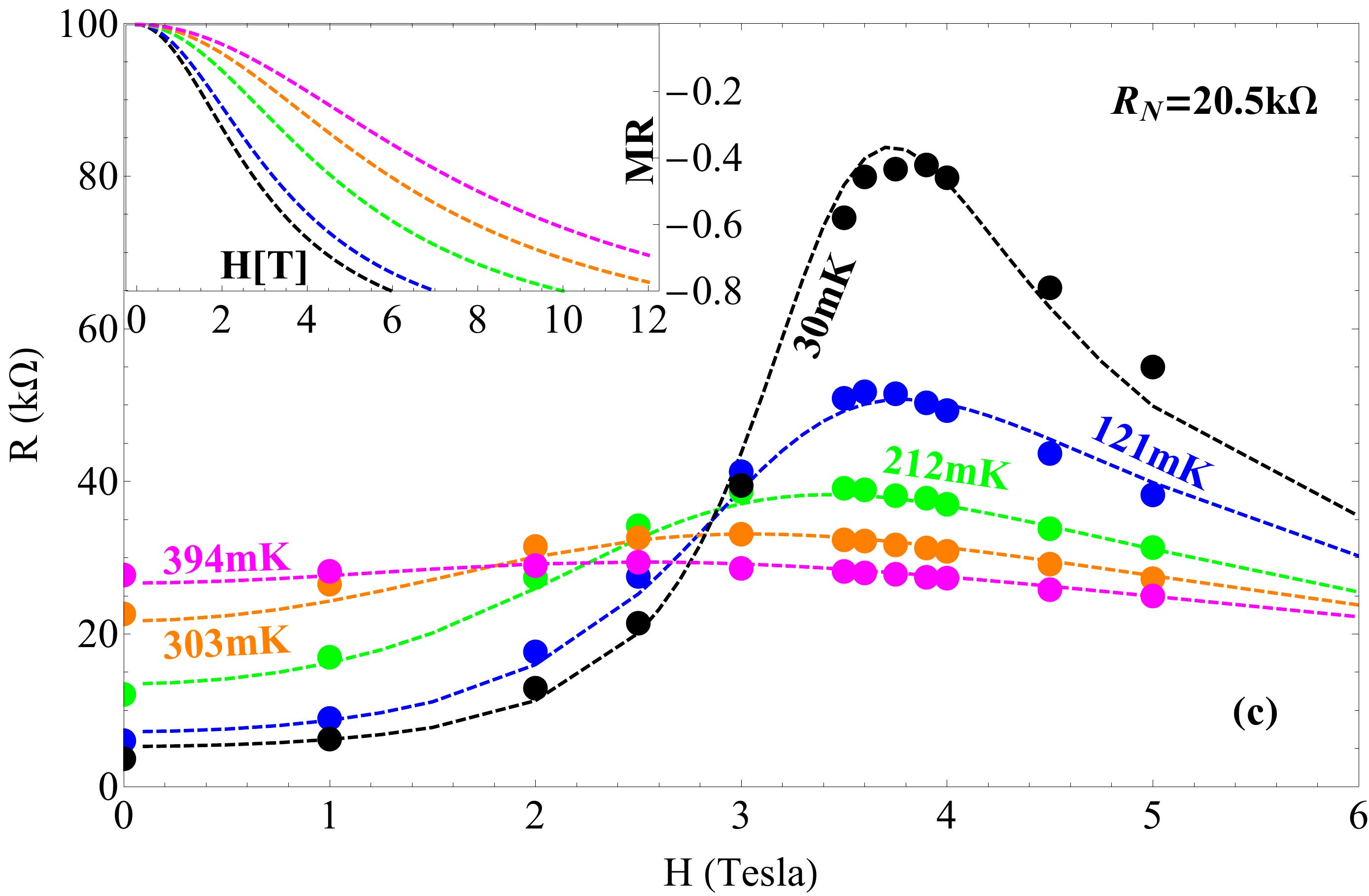}
\caption{ Calculated sheet resistance as a function of field $H$ for $%
R_{N}=7.5$k$\Omega $ (a), $R_{N}=10.5$k$\Omega $ (b), and $R_{N}=20.5$k$%
\Omega $ (c), for different temperatures (dashed lines), plotted together
with the corresponding experimental data (full circles) extracted from Ref. 
\protect \cite{Mograbi19}. The temperature $T$ used in each curve is labeled.
Insets: The normal-state MR curves extracted in the fitting process for each 
$R_{N}$ value. Temperatures follow those of the respective main figure, in
increasing order with the decreasing magnitude of the MR. The corresponding
values of $T_{Q}(T)$ can be found in SM \protect \cite{SuppMat}. Note the
dramatic appearance of resistance at low fields in panel c resulting from
quantum tunneling effect enhanced by the sharply decreasing electronic DOS.}
\end{figure}

The best fitting carrier density $n_{2D}\left( R_{N}\right) $ ($\sim
0.5\times 10^{13}cm^{-2}$) and band effective mass, $m^{\ast }\simeq
1.6m_{e} $, are found in good quantitative agreement with the carrier
density and cyclotron mass, respectively, extracted from SdH oscillations
measurements reported in Re.\cite{KhannaPRL19}. Note that $n_{2D}\left(
R_{N}\right) $ extracted in our fitting is a small fraction of the measured
inverse hall coefficient $e/R_{H}\equiv n_{2D}^{Hall}\left( R_{N}\right) $
reported in Ref.\cite{Mograbi19} ($\sim 10^{14}cm^{-2}$). The situation is
quite similar to that reported for the (001) LaAlO$_{3}$/SrTiO$_{3}$
interface (see \cite{Biscaras12},\cite{Singh18}). The large difference
between the carrier densities extracted from the two methods was attributed 
\cite{Yang16}, \cite{KhannaPRL19} to contributions to transport of at least
two bands with different mobilities, a band contributing minority carriers
with high mobility, dominating the SdH oscillations and superconductivity,
and majority-carriers band with low mobility, which dominate the Hall
resistance.

The key parameter in our theory is the fluctuations-interaction parameter $%
\alpha \left( R_{N}\right) $, which depends on the normal-state sheet
resistance parameter $R_{N}$ \cite{CommentRN}, through $N_{2D}\left(
E_{F}\right) $ (see Eq.\ref{alpha}). For the values of $R_{N}$, presented in
Fig.(3), $\alpha \left( R_{N}\right) $ shows a moderate rise upon increasing 
$R_{N}$ from $7.5$k$\Omega $ to $10.5$k$\Omega $ , and a significantly
larger ascent upon increasing $R_{N}$ from $10.5$k$\Omega $ to $20.5$k$%
\Omega $ (see Fig.1). This has two important consequences seen in Fig.(3)
(see SM \cite{SuppMat} Sec.VIII); a large enhancement of the MSR peak at
high fields, and strong amplification of the quantum tunneling induced
resistance at low fields, which characterizes anomalous metallic behavior 
\cite{KapitulnikRMP19}. The corresponding negative normal state MR curves,
shown in Fig.(3), are seen to exhibit similar enhancements upon increasing $%
R_{N}$, indicating the sharing roles between Cooper-pair (bosonic)
fluctuations and (fermionic) quasi-particles in driving the system to
insulator. The implication with regard to $N_{2D}\left( E_{F}\right) $ is
that, since $N_{2D}\left( E_{F}\right) \propto 1/\alpha \left( R_{N}\right) $%
, its relatively large drop upon down-shifting the Fermi level from $%
E_{F}\left( R_{N}=10.5k\Omega \right) $ to $E_{F}\left( R_{N}=20.5k\Omega
\right) $ reflects electron transfers between bands of considerable
effective mass ratio \cite{KhannaPRL19} (see Fig.(1)), which is, however,
significantly smaller than that calculated for the (001) interface in Refs.%
\cite{JoshuaNcomm12},\cite{DiezPRL15}.

\textit{Discussion}:--The main new message of this letter to the current
understanding of the various SIT phenomena is in proposing the concept of
suppressed DOS by Cooper-pairs formation as a dominant origin of the
insulator side of the SIT. The good quantitative agreement found between the
calculated MSR and the extensive experimental data \cite{Mograbi19},
supports this proposal. The presence of disorder-induced spatial
inhomogeneity, in the form of SC islands, which has been extensively
discussed in the SIT literature \cite{Dubi07}, \cite{Bouadim2011}, \cite%
{GhosalPRL98}, \cite{Vinokur2008}, is reflected in our approach by the
Fourier transform to real space of the fluctuation propagator in the dirty
limit $D\left( q;\widetilde{\varepsilon }_{H}\right) $, which reveals the
underlying structure of mesoscopic SC puddles, whose average localization
length is: $\xi _{H}=\left[ \left( \hbar D/4\pi k_{B}T\right) \eta \left(
H\right) /\widetilde{\varepsilon }_{H}\right] ^{1/2}$ (see SM \cite{SuppMat}
for details).

The emerging physical picture is as follows: Upon increasing the magnetic
field towards the sheet-resistance peak region the 2D SC fluctuations system
breaks into mesoscopic puddles of localized Cooper-pair fluctuations, which
consume much of the unpaired mobile electrons contributing to the normal
state conductivity. The localization arises from opening of an energy gap, $%
\widetilde{\varepsilon }_{H},$ in the fluctuations spectrum and diminishing
their effective stiffness coefficient, $\hbar D\eta \left( H\right) $,
occurring upon increasing magnetic field. In parallel with this localization
process upon increasing field, the AL conductivity decreases more sharply
than the absolute value of the DOS conductivity, so that at the point of
their crossing one observes the onset of the insulating state.

Dynamical quantum tunneling enhances localization and suppresses the density
of SC puddles by enhancing the energy gap $\widetilde{\varepsilon }_{H}$, so
that in the low-fields region, where the AL conductivity is dominant, finite
resistance is generated (even at zero temperature), leading to anomalous
metallic behavior \cite{KapitulnikRMP19}.

Notwithstanding the general nature of the proposed mechanism for SIT, the
role played by spin-orbit scattering in this system is found quite unique:
It strongly mixes relatively heavy electron bands with a lighter electron
band and so sharply suppresses the effective DOS upon down shifting the
chemical potential across a Lifshitz point. The latter effect is associated
with strong enhancement of interaction between Cooper-pair fluctuations,
which at low temperatures, significantly enhances the sheet resistance peak
at high fields and strongly amplifies the low resistance in the low fields
region \cite{KapitulnikRMP19}.

\textbf{ACKNOWLEDGMENTS }

We would like to thank Y. Dagan, E. Maniv and M. Mograbi for useful
discussions and for providing us with their experimental data, which have
been the empirical basis for this paper.


\begin{thebibliography}{99}
\bibitem{CommentRN} Since the gate voltage $V_{g}$ is not uniquely
determined, the measured parameters are labeled by $R_{N}$-the corresponding
sheet resistance measured at zero field and sufficiently high temperature ($%
T=1$K), see Ref.\cite{Mograbi19}
\end{thebibliography}
\end{document}